\documentclass[11pt]{article}
\usepackage{bbm}
\usepackage{amsmath,amssymb,color,epsfig}
\usepackage{graphicx}
\usepackage{subfigure}
\usepackage{setspace}
\usepackage{braket}
\usepackage{multirow}
\usepackage[numbers,sort&compress]{natbib}

\usepackage{amsfonts}

\makeatletter
\@addtoreset{equation}{section}
\makeatother

\newcommand{\be}{\begin{equation}}
\newcommand{\ee}{\end{equation}}
\newcommand{\bea}{\setlength\arraycolsep{2pt} \begin{eqnarray}}
\newcommand{\eea}{\end{eqnarray}}
\newcommand{\nn}{\nonumber}

\newcommand{\bpm}{\begin{pmatrix}}
\newcommand{\epm}{\end{pmatrix}}

\newcommand{\td}{\mathrm{d}}
\newcommand{\te}{\mathrm{e}}

\def\ft#1#2{{\textstyle{\frac{\scriptstyle #1}{\scriptstyle #2} } }}
\def\fft#1#2{{\frac{#1}{#2}}}

\def\0{{\sst{(0)}}}
\def\1{{\sst{(1)}}}
\def\2{{\sst{(2)}}}
\def\3{{\sst{(3)}}}
\def\4{{\sst{(4)}}}
\def\5{{\sst{(5)}}}
\def\6{{\sst{(6)}}}
\def\7{{\sst{(7)}}}
\def\8{{\sst{(8)}}}
\def\sst#1{{\scriptscriptstyle #1}}

\begin{document}


\begin{center}
{\large {\bf Extremal Charged Black Holes and Superradiantly Unstable Quasinormal Modes}}

\vspace{10pt}
Zhan-Feng Mai, Run-Qiu Yang and H. L\"{u}

\vspace{10pt}

{\it Center for Joint Quantum Studies and Department of Physics\\ School of Science, Tianjin University,\\ Yaguan Road 135, Jinnan District, Tianjin 300350, China}

\vspace{40pt}

\underline{ABSTRACT}
\end{center}

It was recently shown that the extremal charged black holes in STU supergravity suffer from superradiant instability owing to the existence of the unstable (low-frequency) quasibound states associated with a charged massive scalar field.  In this paper, we show numerically that for some charge configurations, these black holes can also excite the (higher-frequency) superradiantly unstable quasinormal modes. We find empirically that the unstable modes are closely related to having a volcano-shaped effective potential in the Schr\"odinger-like wave equation.

\vfill {\footnotesize  zhanfeng.mai@gmail.com \ \ \ aqiu@tju.edu.cn \ \ \ mrhonglu@gmail.com}

\thispagestyle{empty}

\pagebreak



\newpage

\section{Introduction}

Whether a black hole is stable or not under a small perturbation is one of its most important properties. It not only has consequences in astronomical observations, but also in the study of fundamental physics, since extremal charged and/or rotating black holes are sometimes treated as fundamental particles in strings or M-theory. At the linearized level, the Schwarzschild black hole has been proven to be stable against the scalar, vector and tensor perturbations \cite{Chandrasekhar:1975zza,Iyer:1986np,Iyer:1986nq}. It was discovered that superradiant effect can occur for charged scalar perturbations in the background of charged and/or rotating black holes such that energy can be extracted from these black holes when the scalar satisfies the suitable ingoing boundary condition on the horizon that is also superradiant \cite{Starobinsky:1973aij,Damour:1976kh,Brito:2015oca}. In the frequency domain where our paper is focused on, there exist two distinct modes that could be responsible for instability, both of which must satisfy the ingoing boundary condition on the black hole horizon. At asymptotic infinity, the scalar equation reduces to the Klein-Gordon equation in the Minkowski space.  When the frequency $\omega$ is smaller than the scalar mass, the spatial sector either falls off or diverges exponentially.  In this case, we must impose the exponential-falloff condition, leading to a potential quasibound state (QBS). On the other hand, when $\omega$ is bigger than the mass, the spatial sector oscillates and we must choose the purely outgoing boundary condition to ensure that there is no external source.  This leads to a potential quasinormal mode (QNM). This classification provides a new venue of studying the black hole stability, since it is natural to ask whether a superradiantly unstable QNM or QBS, which has no external energy source, can cause instability, at least at the linear level, by simply extracting energy from the black hole. It can be shown \cite{Mai:2021yny} that the eigenfrequencies associated with the time-dependent factor $e^{-i \omega t}$ of either QBS's or QNMs in a stationary black hole background are necessarily complex, namely
\be
\omega=\omega_r + i \omega_i\,.\label{complexomega}
\ee
Thus the mode with negative $\omega_i$ will decay exponentially with time, indicating that the black hole is stable. However, when $\omega_i$ is positive, the mode will exponentially grow with time, and the spacetime is unstable, at least at the linear level.

For a charged massive scalar field $\Phi$ of mass $m_p$ and charge $q$ in the asymptotically-flat black hole background, QBS's can arise at low frequency $(\omega_r < m_p)$ and QNMs arise only at higher frequency $(\omega_r > m_p)$. The Kerr black hole and Kerr-Newman black hole were shown to suffer from superradiant instability owing to the existence of the unstable QBS's. The Reissner-Nordstr\"om (RN) black hole, on the other hand, is stable under such a perturbation. The non-existence of unstable QBS's in the RN black hole was established in \cite{Furuhashi:2004jk,Hod:2012wmy,Huang:2015jza,Hod:2015hza,Huang:2022nzm}. However, it was recently shown  \cite{Mai:2021yny} that multicharge extremal black holes in STU supergravity \cite{Duff:1995sm} can suffer from such low-frequency superradiant instability as long as not all the charges of the black hole are equal (When all the charges are equal, the black hole becomes the RN black hole.)

The story of QNMs is similar; the QNMs constructed in the background in the Kerr, Kerr-Newman or RN black holes are all stable \cite{Leaver:1985ax,Leaver:1990zz,Natario:2004jd,Konoplya:2006br,Percival:2020skc}. (The situation may change in the asymptotic (anti-)de Sitter spacetime ((A)dS) \cite{Konoplya:2014lha,Zhu:2014sya}, which we do not consider in this paper.) Inspired by the QBS result of \cite{Mai:2021yny}, we construct the QNMs in extremal black holes that can be embedded in STU supergravity. However, it should be emphasized that the existence of the unstable QBS's in \cite{Mai:2021yny} does not necessarily imply the existence of unstable QNMs. They belong to very difference classes of solutions, owing to the fact that the eigenfrequency $\omega$ enters the charged scalar equation nontrivially and its value can alter the shape of effective potential. For example, although the Kerr black hole admits unstable QBS's, the QNMs on the other hand are all stable \cite{Leaver:1985ax,Dolan:2007mj}. Instabilities in the five-dimensional rapidly rotating black hole perturbed by nonaxisymmetric spin-2 tensor were found in \cite{Shibata:2009ad}, but the analysis was done in the time domain. In fact, even though the instabilities of various black holes have been well studied, there is hitherto no known example of unstable QNMs (of frequency domain) in literatures. In this paper, we find that when the charges of the STU black holes are sufficiently different, the multicharge black holes can also excite unstable QNMs, therefore establishing that superradiant instability by unstable QNMs can occur among the static charged black holes.

The paper is organized as follows. In Sec.~\ref{modset}, we briefly introduce the extremal charged black hole background and study the linearized perturbation of a charged massive scalar field. We give the boundary conditions that lead to the QNMs. In Sec.~\ref{cond}, we discuss the necessary superradiant conditions for the unstable QNMs. We also analyse the structure of the effective potentials in the Schr\"odinger-like equation that may be related to the unstable QNMs. In Sec.~\ref{result}, we present our explicit examples of numerical results of the unstable QNMs. We conclude the paper in Sec.~\ref{conclu}.

\section{Quasinormal modes of extremal black holes}\label{modset}

We begin with a class of string-theory inspired Einstein-Maxwell-Dilaton (EMD) theories involving two Maxwell fields $A_1$ and $A_2$, which have non-minimal couplings with a dilatonic scalar $\phi$. The Lagrangian is
\begin{equation}\label{action}
{\cal L}= \sqrt{-g} \left(R - \frac{1}{2}(\partial \phi)^2-\frac{1}{4}\te^{\alpha_1 \phi}F_1^2-\frac{1}{4}\te^{\alpha_2 \phi}F_2^2  \right),
\end{equation}
where $F_i= \td A_i$ ($i=1,2$) are the field strengths. The two dilatonic coupling constants $(\alpha_1, \alpha_2)$ satisfy \cite{Lu:2013eoa}
\begin{equation}\label{N1N2}
\alpha_1 \alpha_2 = -1 \, , \quad N_1 \alpha_1 + N_2 \alpha_2 = 0 \, ,  \quad N_1 + N_2 = 4 \, .
\end{equation}
For integer values, i.e.~$(N_1,N_2)=(2,2)$ or $(1,3)$, the theory can be embedded into the STU supergravity model \cite{Duff:1995sm}. In particular, the theory with $(N_1,N_2)=(2,2)$ can be embedded in ${\cal N}=2$, $D=4$ supergravity coupled to a vector super-multiplet.  The Lagrangian \eqref{action} with \eqref{N1N2} admits exact solutions of electrically charged black holes \cite{Lu:2013eoa}.  In the extremal limit, they are
\begin{eqnarray}\label{exbg}
&&\td s^2=-\left(H_1^{N_1}H_2^{N_2} \right)^{-\frac{1}{2}}\td t^2 +(H_1^{N_1} H_2^{N_2})^{\frac{1}{2}}\left(\td r^2 + r^2 \td \Omega^2_{2} \right)\,, \cr
&& A_i = \sqrt{N_i} \left(H_i^{-1} -1  \right) \td t \,,\qquad \phi =\sum_{i=1}^2 \frac{1}{2}\alpha_i N_i \log H_i\, ,\qquad H_i =1+ \fft{4Q_i}{\sqrt{N_i}\,r}\,.
\end{eqnarray}
The solution is specified by two integration constants, namely the electric charges $(Q_1,Q_2)$ associated with the Maxwell fields $(A_1, A_2)$.  The black hole mass is
\be
M=\sqrt{N_1} Q_1 + \sqrt{N_2} Q_2\,.\label{masscharge}
\ee
The solution interpolates between the asymptotic Minkowski spacetime at $r\rightarrow \infty$ and the AdS$_2\times S^2$ near-horizon geometry at $r\rightarrow 0$. The spacetime curvature singularities are located at some negative $r$ where either $H_1$ or $H_2$ vanishes. These singularities are shielded by the horizon at $r=0$, where $g_{tt}$ vanishes.  The near-horizon geometry is regular and AdS$_2\times S^2$, as in the case of the extremal RN black hole. The Riemann tensor square near the horizon is
\begin{equation}\label{R2}
\left. R^{\mu\nu\rho\sigma}R_{\mu\nu\rho\sigma}\right|_{r \to 0} = \ft{1}{32} N_1^{N_1/2}N_2^{N_2/2}Q_1^{-N1}Q_2^{-N_2} + {\cal O}(r) \,.
\end{equation}
It is easy to observe when one of the charges vanishes, the horizon becomes singular; therefore, we shall consider $Q_1Q_2\ne0$ only.  When $\frac{Q_1}{\sqrt{N_1}} = \frac{Q_2}{\sqrt{N_2}} $, these solutions reduce to the extremal RN black hole.

As in our previous paper \cite{Mai:2021yny}, we consider an additional massive complex scalar $\Phi$ that is charged under both Maxwell fields. Its linearized perturbation can be described by the charged Klein-Gordon (KG) equation
\begin{equation}\label{kg1}
\left(g^{\mu\nu} D_\mu D_\nu - m_p^2\right) \Phi=0 \, , \quad \quad D_{\mu} := \nabla_\mu - i  q_1  A_{1 \mu} - i  q_2  A_{2\mu} \, ,
\end{equation}
where $(m_p, q_i)$ denote the scalar's fundamental mass and electric charges. It should be clarified that the purpose of the paper is to establish whether static charged black holes can have unstable QNMs; therefore, we add the above simplest linear perturbation even though it is likely to break the supersymmetry of the theory. Whether the supersymmetric extremal charged black hole in STU supergravity is stable or not should be addressed in the context of supergravity only.

In the black hole background \eqref{exbg}, the general solution of the KG equation can be expressed as linear superpositions in the frequency domain
\begin{equation}\label{anz}
\Phi = \te^{- i \omega t} R(r) {\rm Y}_{\ell m}\left(\theta, \varphi\right),
\end{equation}
where $Y_{\ell,m}$'s are the standard spherical harmonics. The KG equation \eqref{kg1} then reduces to the radial equation
\bea\label{radeq}
&&-r^2 \frac{\td }{ \td r}\left( r^2 \frac{\td R}{\td r}  \right) + U(r) R = 0 \,,\\
U&=&-r^4 H_1^{N_1} H_2^{N_2}\left(\omega-\sum_{i=1}^2 \fft{4q_i Q_i}{r H_i} \right)^2+m_p^2 r^4 H_1^{N_1/2}H_2^{N_2/2}+\ell(\ell+1)r^2 \,.\nn
\eea
The radial equation cannot be solved in general. However, in both the horizon ($r=0$) and asymptotic ($r\rightarrow \infty)$ regions, the equation is solvable. Specifically, we have
\bea
r\rightarrow 0:&&\qquad U = -\lambda^2 (\omega - \omega_c)^2 + {\cal O}(r)\,,\qquad \lambda=\Big(\fft{4Q_1}{\sqrt{N_1}}\Big)^{\fft{N_1}{2}} \Big(\fft{4Q_2}{\sqrt{N_2}}\Big)^{\fft{N_2}{2}},\nn\\
r\rightarrow \infty:&&\qquad U = -k^2\, r^4 \Big(1 + {\cal O}(\fft{1}{r^2})\Big)\,,\qquad
k=\sqrt{\omega^2-m_p^2}\,,
\eea
where $\omega_c$ is the critical value of frequency that is linearly dependent on the scalar's fundamental charges
\begin{equation}
\omega_c = \sqrt{N_1}q_1 + \sqrt{N_2} q_2 .\label{omegac}
\end{equation}
It is intriguing to note that the scalar's fundamental charges and mass provide critical values of frequency to the characteristic behavior of the effective potential on the horizon and at asymptotic infinity respectively. We therefore have
\begin{equation}\label{asy1}
R(r)|_{r \to 0} = \te^{i\fft{\lambda\,(\omega - \omega_c)}{r}} \, , \qquad  R(r)|_{r \to \infty} = {\cal T} \frac{\te^{i k r} }{r}\,.
\end{equation}
Note that the solution allows to take $\pm \lambda$ and $\pm k$, and we choose the positive sign for $\lambda$ so that wave is ingoing on the horizon, and the positive sign for $k$ so that it is outgoing at asymptotic infinity. Without loss of generality, we set the overall coefficient to be one on the horizon, and consequently, the coefficient $\cal T$ describes the transmission rate to the asymptotic infinity. In the above discussion, we have implicitly assumed
\begin{equation}\label{cond1}
\omega > m_p \,.
\end{equation}
In fact, the asymptotic behavior of $R(r)$ depends crucially on the value of $\omega$.  For low-frequency modes $(\omega<m_p)$, $k$ becomes imaginary and the function $R(r)$ falls off exponentially, giving rise to QBS's, which were studied in detail in our previous paper \cite{Mai:2021yny}.  Here, we focus instead on the high-frequency modes with \eqref{cond1}, such that the asymptotic $R(r)$ is wavelike. We shall  construct the QNMs whose boundary condition is outgoing asymptotically, i.e.~it is specified by \eqref{asy1}. In \cite{Mai:2021yny}, we proved that the frequency {\it must} be complex, as shown in Eq.~\eqref{complexomega}. Typically in literature, the QNMs have negative $\omega_i$, and they will decay exponentially with time. The QNMs with positive $\omega_i$, on the other hand, are unstable since they will grow exponentially with time. In this paper, we investigate whether unstable QNMs can arise in black hole spacetime. Since there is hitherto no such an example, it is instructive first to study some necessary conditions that the unstable QNMs could arise.

\section{Searching for unstable quasinormal modes}\label{cond}

\subsection{Necessary condition on frequency}

Intuitively, in order to have unstable QNMs that grow exponentially with time, it is necessary that they are superradiant in that energy can be extracted from the black hole, since there is no energy pumping in from the asymptotic region. The analysis on both the energy and charge currents turned out to be quite effective in obtaining the superradiant conditions for the unstable QBS's \cite{Mai:2021yny}. We therefore adopt the same method to analyse the QNMs.  The conserved energy current $J_{\rm E}^\mu$ and charge currents $I_{\rm Q}^{\mu}$ are
\begin{equation}
J^\mu_{\rm E} = T^\mu{}_\nu \left(\frac{\partial}{\partial t} \right)^\nu\,, \qquad i I_Q^\mu = \Phi^\dag D^\mu \Phi - \Phi (D^\mu \Phi)^{\dag} \, ,
\end{equation}
where the energy momentum tensor $T_{\mu\nu}$ is \cite{DiMenza:2014vpa}
\begin{equation}\label{enercurr}
T_{\mu\nu} = \sum_{j=r,i}\left( 2\nabla_\mu \Phi_j \nabla_\nu \Phi_j  -  g_{\mu\nu}\left(\nabla^\rho \Phi_j \nabla_\rho \Phi_j +  \left(\sum^2_{k=1}  q_k  A^\rho_k \right)^2  \Phi_j^2 \right) - m_p^2\Phi^2_j\right) \, ,
\end{equation}
and $\Phi_r = \frac{1}{2} \left( \Phi + \Phi^{\dag} \right)$ and $\Phi_i = \frac{1}{2 i} (\Phi - \Phi^{\dag})$. Both currents satisfy their conservation laws, namely
\begin{equation}\label{cons}
\nabla_\mu J^\mu _{\rm E} = 0 \, ,    \qquad \nabla_\mu I_Q^\mu = 0 \, .
\end{equation}
For modes with complex frequencies, one can read off the growth rates of the total energy and charge outside of the horizon from $J^{\mu}_{\rm E}$ and $I_{\rm Q}^\mu$ respectively, i.e.
\begin{equation}\label{dE1}
\frac{\partial  E}{\partial  t} = 2 \omega_ i E  \, , \quad \frac{\partial Q}{\partial t} = 2 \omega_i Q\, , \quad  \text{where} \quad E \equiv \int_{V} \sqrt{-g}J^0_{\rm E} \, ,  \quad Q \equiv \int_{V} \sqrt{-g} I_Q^0 \, .
\end{equation}
Note that both growth rates have the same sign dependence on the imaginary part of the frequency: the state is stable when its $\omega_i< 0$, and unstable when $\omega_i>0$.

We now derive a necessary superradiant condition for the unstable QNMs. The conservation law \eqref{cons} implies that
\begin{equation}\label{Gauss}
\left. \frac{\partial }{\partial t} \int_V J^{0}_{\rm E}  =\frac{4 \pi}{2\ell+1}\, J^r_{\rm E} \right|^{r \to \infty}_{r \to 0} \, , \qquad
\left. \frac{\partial Q}{\partial t} = \frac{4 \pi}{2\ell+1}\, \int_V  I_Q^r \right|^{r \to \infty}_{r \to 0} \,.
\end{equation}
We see that the results depend on only the boundary conditions, which we specified in Eq.~\eqref{asy1}. It follows from Eqs.~\eqref{dE1} and \eqref{Gauss} that we have
\begin{eqnarray}
&& 2E \omega_i \varpropto - \omega_r(\omega_r- \omega_c) - |{\cal T}|^2\omega_r \sqrt{\omega_r^2 - m_p^2} + {\cal O } (\omega_i^2 ) \, , \cr
&& \cr
&& 2Q \omega_i \varpropto - (\omega_r- \omega_c) - |{\cal T}|^2 \sqrt{\omega_r^2 - m_p^2} + {\cal O } (\omega_i^2 ) \,,
\end{eqnarray}
where we have stripped off the overall positive numerical numbers for simplicity. We therefore conclude that a necessary condition for having $\omega_i > 0$ is
\begin{equation}\label{supcond}
\omega_r < \omega_c \, .
\end{equation}
The small $\omega_i$ expansion in the analysis is consistent with our numerical result that $\omega_i/\omega\ll 1$. It is worth commenting here that the term associated with the transmission coefficient ${\cal T}$ does not exist in the QBS's; its negative sign represents that both the energy and charges can only flow out to infinity. The physical origin of the instability becomes clear. When $\omega_r<\omega_c$, the superradiant effect takes place, and the energy and charges are extracted from the black hole.  Some of these energy and charges outflow to infinity. When the extraction overpowers the outflow, the trapped energy and charges will grow exponentially.

Since the QNMs require that $\omega_r > m_p$, the condition Eq.~\eqref{supcond} gives a constrain related to the fundamental mass and charge of the scalar,
\begin{equation}
m_p < \sqrt{N_1} q_1 + \sqrt{N_2} q_2 \,.\label{upmp}
\end{equation}
In other words, unstable QNMs arise only for the larger charge/mass ratio of the scalar field.

It should be pointed out that what we have presented are the necessary conditions only. The system may be stable after all, which is indeed the case for most known black holes. This is because the \eqref{supcond} can be easily violated for two related reasons. One is that the transmission coefficient ${\cal T}$ is still to be determined and it can be large.  The second is that the complex frequency spectrum is not continuous, but discretized.  In other words, the discrete set of the frequencies $\omega_{\rm QNM}$ of the QNMs simply do not satisfy \eqref{supcond}. For example, there is no such unstable QNMs in the RN black hole. Nevertheless, the necessary conditions we obtained above can guide us in the search of the unstable QNMs.

\subsection{Effective potentials}

It is clear that ultimately the properties of the QNMs are determined by the function $U(r)$ given in Eq.~\eqref{radeq}. It is instructive to cast the radial equation Eq.~\eqref{radeq} into the Sch\"{o}dinger-like form in the tortoise coordinate. We first define a new radial function $\tilde R$ and the tortoise coordinate $y$ as
\begin{equation}
\frac{\td y}{\td r} = H_1^{N_1/2} H_2^{N_2/2}\, , \quad  \tilde R = H_1^{-N_1/4}H_2^{-N_2 / 4} \frac{R}{r} \, .
\end{equation}
The radial equation Eq.~\eqref{radeq} becomes the Schr\"{o}dinger-like equation as
\begin{equation}
-\frac{\td^2 \tilde R}{\td y^2 } + \bar{U}(y) \tilde R = \omega^2 \tilde R\, , \qquad \bar U = 2 \omega Q(y)-Q(y)^2 + V(y) \, ,
\end{equation}
where
\bea
&&Q(y)=\sum^2_{i=1}\sqrt{N_i}\, q_i(1-H^{-1}_i)\, ,\qquad \quad H = H_1^{N_1} H_2^{N_2}\, ,\nn\\
&& V(y)= \frac{\ell (\ell + 1)}{r^2 H}+ \frac{m_p^2}{\sqrt{ H}} - \frac{5  H'^2}{16  H^3} + \frac{ H''}{4 H^2}\,.
\eea
Here a prime denotes a derivative with respect to $r$. The unstable QBS's of $\omega < m_p$ were constructed and analysed in \cite{Mai:2021yny}. They can arise when $\sqrt{N_1}Q_2 \ne \sqrt{N_2} Q_1$. There was a clear guideline in finding QBS's, namely the effective potential $\bar U$ has a deep enough potential well that is sandwiched by two peaks. When $\sqrt{N_1}Q_2 = \sqrt{N_2} Q_1$, the background reduces to the RN black hole, and the double peak potential does not exist. Unfortunately, this simple picture of the effective potentials does not carry over when $\omega > m_p$.  This is because unlike the standard Schr\"odinger equation, the effective potential here depends also on the eigenfrequency $\omega$.  Even we hold all the rest parameters fixed, changing the value of $\omega$ can also change the shape of the effective potential.  For this reason, as was pointed in the introduction, the existence of unstable QBS's does not necessarily imply the existence of unstable QNMs. Nevertheless, by adjusting various parameters, we find that two types of effective potentials emerge: one with double peaks and the other with single peak only.  Here we present a concrete example of $(N_1, N_2)= (2,2)$.

\begin{figure}[hbtp]
  \centering
  \includegraphics[width=0.45\textwidth]{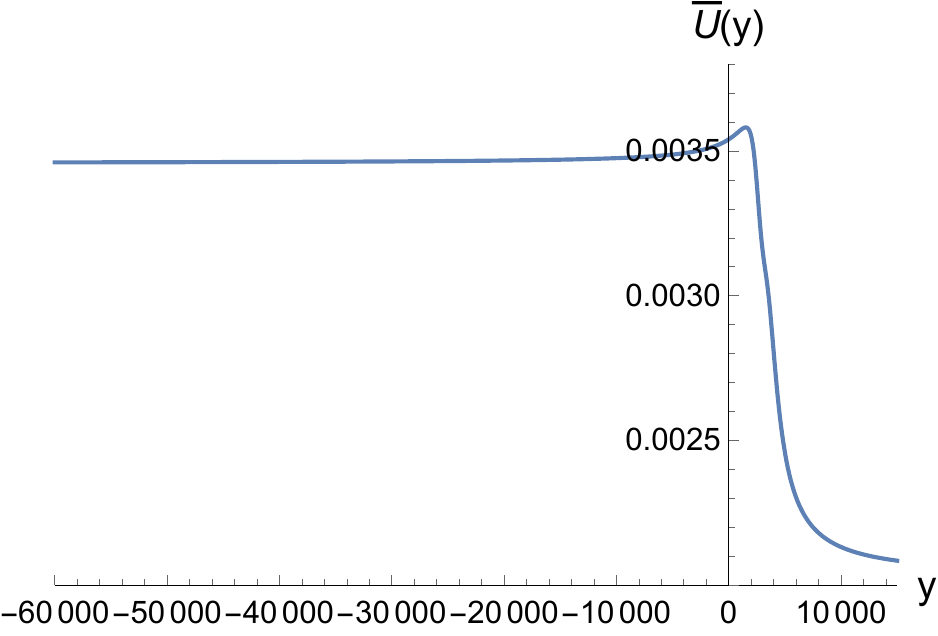}\ \
  \includegraphics[width=0.45\textwidth]{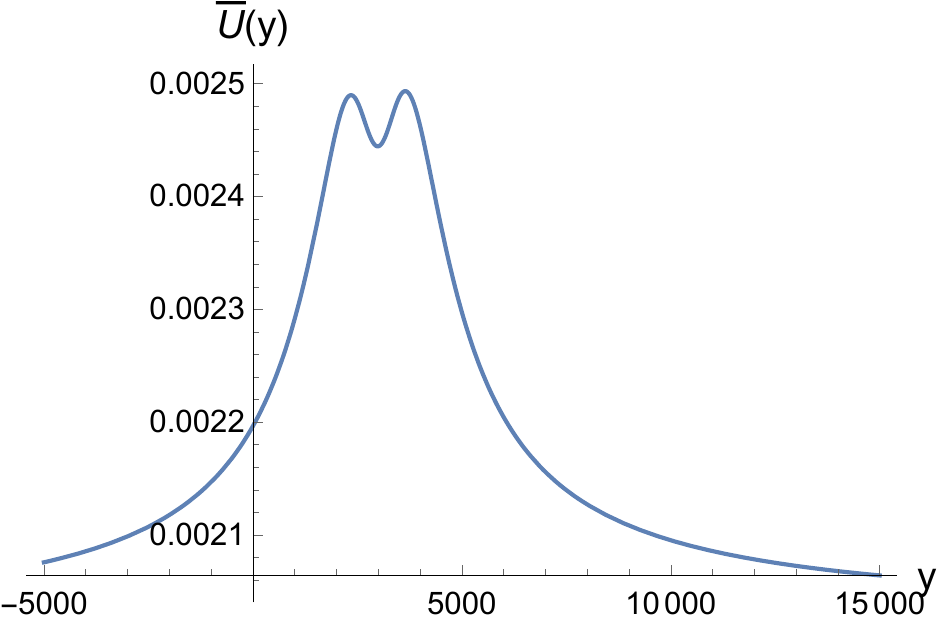}
 \caption{\small There are two types of the effective potentials (in the tortoise coordinate). For example, we fix parameters $(m_p, q_1,q_2,\ell,Q_1,Q_2)$ =(0.045, 0.025, 0.025, 1, 1, 100) and $(N_1,N_2)=(2,2)$.  The left panel has $\omega=0.0598$ and the potential has a single peak. It is like a step function with a bump on the top. The right has $\omega=0.0496$ and the potential has double peaks. It is the latter volcano-shaped potential that gives rise to unstable QNMs.}\label{effp}
\end{figure}

In Fig.~\ref{effp}, we plot the effective potential in tortoise coordinate $y$. we have chosen $(m_p, q_1,q_2,\ell,Q_1,Q_2)=(0.045, 0.025, 0.025, 1, 1, 100)$. When $\omega=0.0598$, the single peak potential resembles a step function with a bump on the top of the cliff. The shape of the potential changes dramatically when we set $\omega=0.0496$ and it has now two peaks. In other words, the potential has three local extrema: the two maxima $0.002490$ and $0.002493$  are located in $y = 1691$ and $y = 2988$ and one minimum $0.002445$ is located in $y = 2330$. Note that the potential well is rather shallow and shaped like a volcano; it cannot bound any state. As we pointed it out earlier, $\omega$ of QNMs must be complex; however, since empirically $\omega_i/\omega_r\ll 1$, we use, a priori, only the real $\omega$ to study the shape of the effective potential.

We find that typically the effective potentials have a single peak and there is no unstable QNM in this case.  For the black holes we considered in this paper, when we set $Q_2/Q_1$ to be sufficiently large, double peaks can arise, but these potentials are not typical and we have to adjust parameters carefully to find them; a small deviation of the parameter can easily destroy one of the two peaks. However, as we shall illustrate in the next section with numerical methods, the volcano-shaped double-peak potential can lead to unstable QNMs.

\section{Numerical results}\label{result}

In this section, we present our numerical results of unstable QNMs.  We begin by
briefly introducing the numerical method, which we perform in the $r$ coordinate.
Since the radial function $R(r)$ can be solved exactly at $r=0$ and $r\rightarrow \infty$, we can perform the power series expansion for both the small and large $r$, anchored by the boundary condition \eqref{asy1}, namely
\bea
R \left( r \to 0\right) &\sim& \left. \te^{i\frac{\chi_1 (\omega - \omega_c)}{r}}r^{-2i \chi_2}\sum^{n_1}_{i=0}r^i h_i \right|_{r = \epsilon_c } \,,\nn\\
R \left( r \to \infty  \right) &\sim&  \left. \te^{i \sqrt{\omega^2-m_p^2} r} r^{\chi_3} \sum^{n_2}_{i=0} \frac{g_i}{r^i}  \right|_{r = r_c}\,,\label{powerseries}
\eea
where $(n_1, n_2)$ denote the expansion order and the coefficients $(\chi_1,\chi_2,\chi_3, h_i, g_i)$ can be analytically solved by the radial equation order by order in terms of $r$. Due to the linearity of the radial equation, we set $h_0 = g_0 = 1$ without loss of generality. Since $r = 0$ and $r = \infty$ are both singularities of the radial equation, they cannot be approached by the numerical method, we denote $\epsilon_c$ and $r_c$ as the numerically cutoffs for the horizon and asymptotic infinity respectively.

By using the Range-Kutta method, we perform numerical integration from $\epsilon_c$ to the midpoint $r_i$, obtaining a numerical solution $R_1$, while from $r_c$ to $r_i$, obtaining another numerical solution $R_2$. $R_1$ and $R_2$ should match at $r_i$ if they describe the same solution, we thus require the Wronskian of $R_1, R_2$ at $r_i$ vanish, namely
\begin{equation}
W(R_1,R_2)= \left. \frac{R_1 R'_2 - R_2 R'_1}{|R_1| |R_2|} \right|_{r=r_i} = 0 \, .
\end{equation}
%
The subtleties of improving numerical accuracies were discussed in detail in \cite{Mai:2021yny}. Here we shall simply present our numerical findings.

\subsection{An example of unstable QNM}

\begin{figure}[hbtp]
  \centering
  \includegraphics[width=0.49\textwidth]{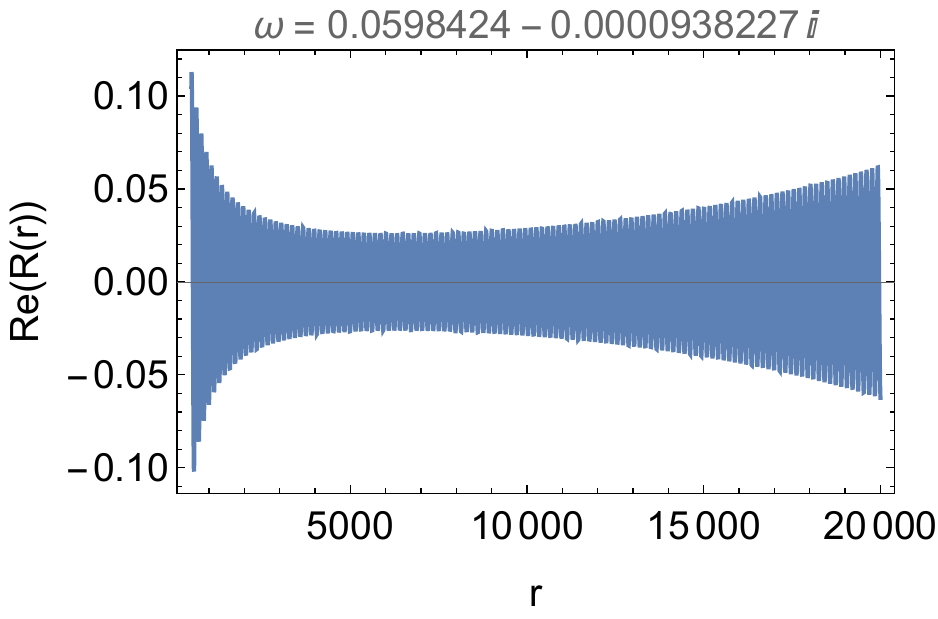}\ \
  \includegraphics[width=0.49\textwidth]{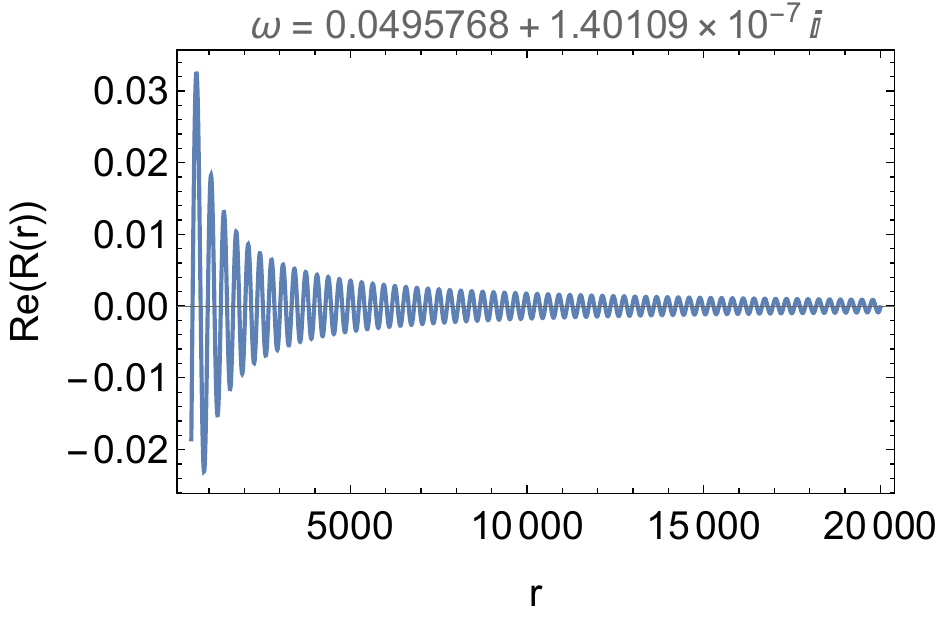}
 \caption{\small For the same fixed parameters $(m_p, q_1,q_2,\ell,Q_1,Q_2)$=(0.045, 0.025, 0.025, 1, 1, 100), both stable (left) and unstable (right) normal modes can arise. Here $\omega_r$ is smaller for the unstable QNM.}\label{plotR}
\end{figure}

We fix the parameters $(m_p, q_1,q_2,\ell,Q_1,Q_2)$=(0.045, 0.025, 0.025, 1, 1, 100). We search for the QNMs whose real frequencies are in the vicinity of either 0.0598 or 0.0496. The corresponding effective potentials were plotted in Fig.~\ref{effp}.  We find that the single-peak potential gives rise to a stable QNM, whilst the volcano-shaped double-peak potential gives rise to an unstable one.  The corresponding complex frequencies are
\begin{equation}
{\rm Stable:}\quad \omega = 0.0598 - 9.38 \times 10^{-5} i \,;\qquad {\rm Unstable:}\quad \omega=0.0496 + 1.40 \times 10^{-7}i \,.
\end{equation}
In Fig.~\ref{plotR}, we plot the real part of the radial function $R(r)$ for both the stable and unstable QNMs. As we can see from the the boundary condition Eq.~\eqref{asy1}, for the stable QNM that was common in literature, the asymptotic radial function $R(r)$ grow exponentially at large $r$, due to the $\te^{-\omega_i r}$ factor. On the other hand, for the unstable QNM, where $\omega_i$ is positive, $R(r)$ falls off exponentially at large $r$.

\begin{figure}[hbtp]
  \centering
  \includegraphics[width=0.4\textwidth]{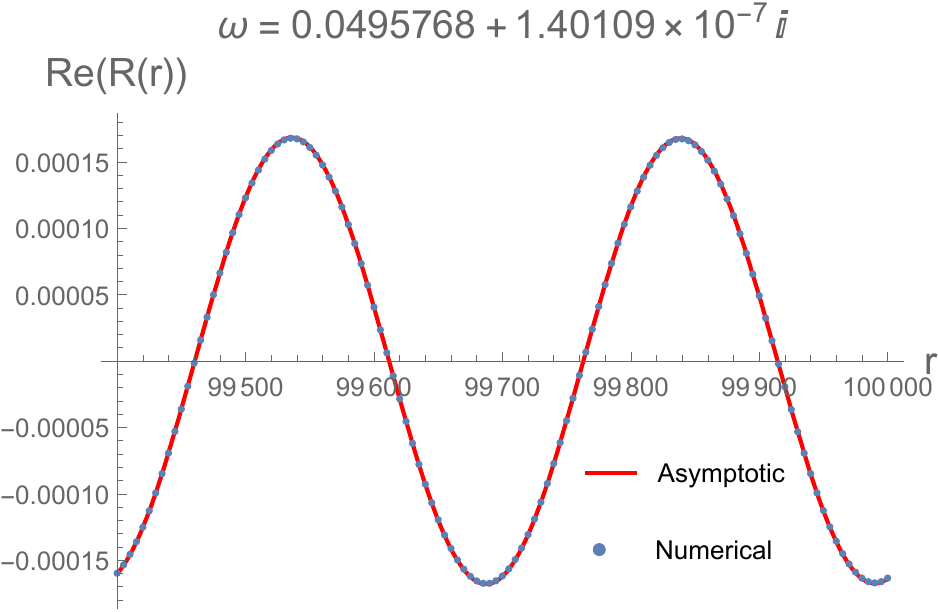}~~
  \includegraphics[width=0.4\textwidth]{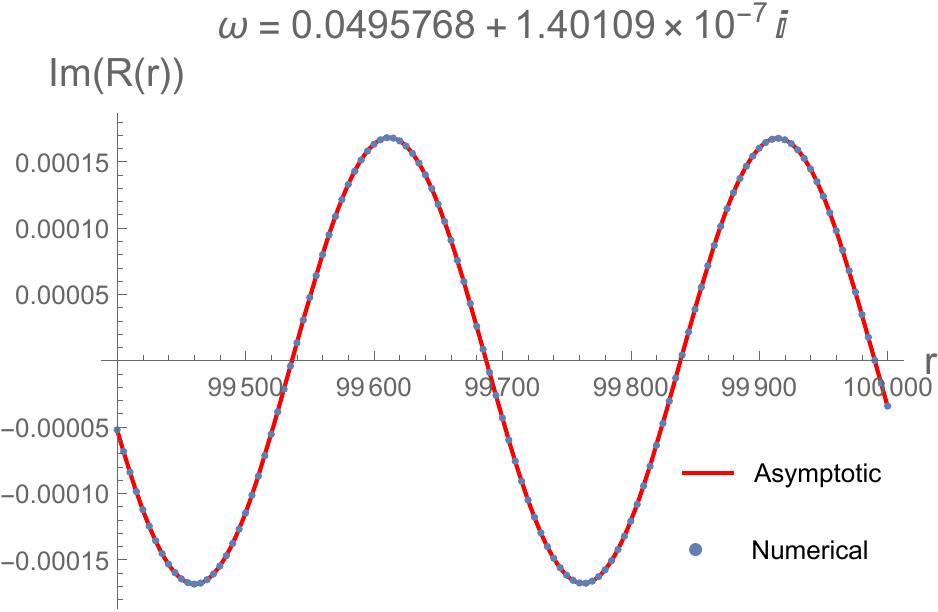}\\

  \includegraphics[width=0.4\textwidth]{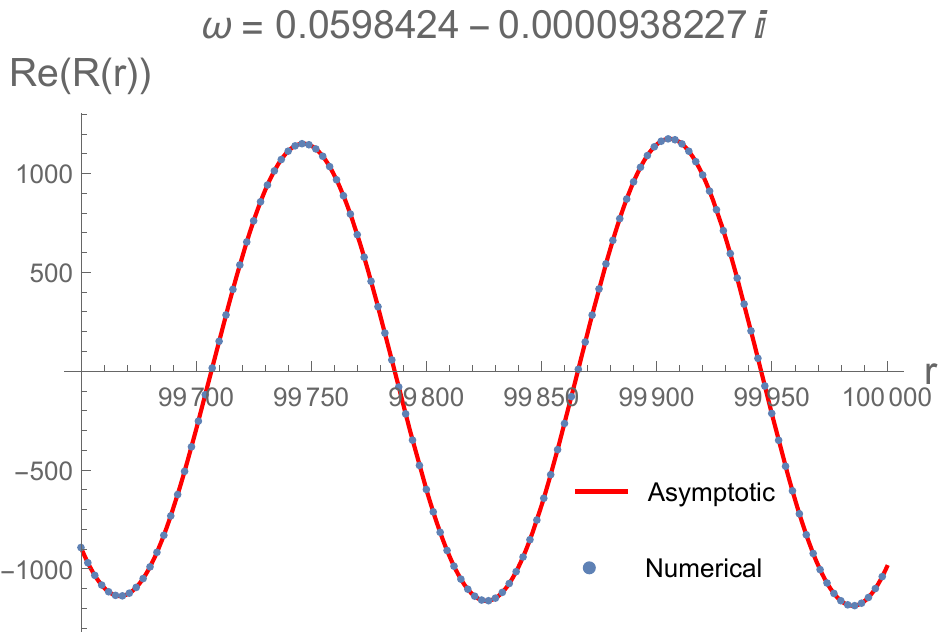}~~
  \includegraphics[width=0.4\textwidth]{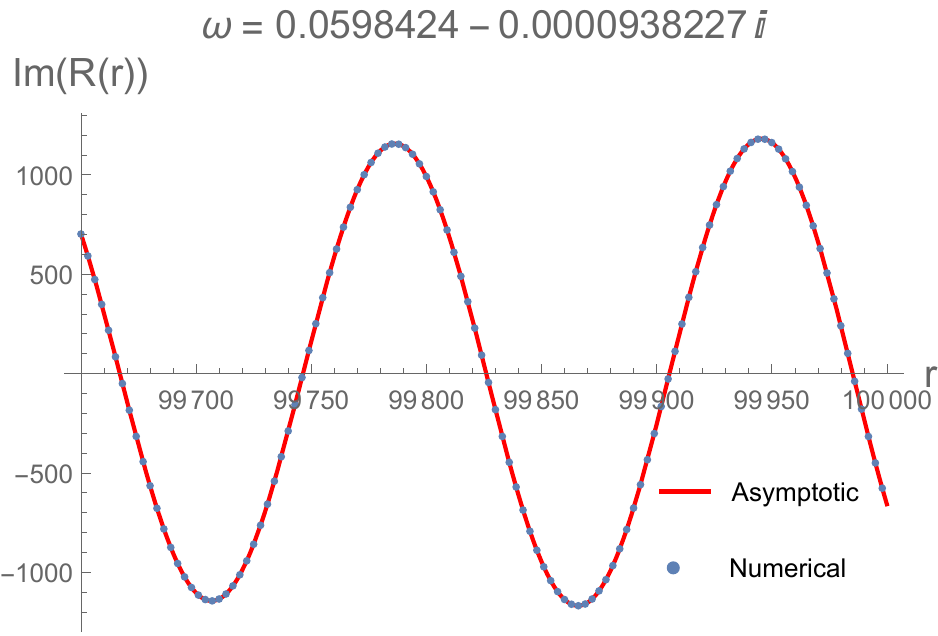}
 \caption{\small These plots illustrate how well the numerical data (dots) fit both the real and imaginary parts of the analytical power series expansion of $R$ at large $r$, for both the stable and unstable QNMs.} \label{asymplot}
\end{figure}

Having obtained the first numerical example of unstable QNM in literature, it is necessary to verify that our numerical construction is sufficiently accurate to be trustworthy, since the integration covers from $\epsilon_c$ to $r_c$.  We have chosen sufficiently small $\epsilon_c\sim 10^{-2}$ to ensure that the power series is well convergent. By adjusting the numerical accuracy, we can further decreasing $\epsilon_c$ to be $10^{-3}$ and $10^{-4}$, etc., we find that our numerical results are convergent, with the relative errors less than $10^{-12}$. This ensures that the boundary condition near the horizon is indeed well imposed.  However, in order to save time in the numerical calculation, our $r_c$ was chosen to be around 400.  It is thus necessary, after obtaining the solution, to integrate out the solution further to larger $r$ and compare the numerical solution to the large-$r$ power series expansion that was obtained analytically. In other words, we would like to compare our numerical results to the asymptotic expression
\begin{equation}\label{Ratlarger}
R_{\rm asy} \left( r \to \infty  \right) \sim  \left. \te^{i \sqrt{\omega^2-m_p^2} r} r^{\chi_3} \sum^{n_2}_{i=0} \frac{g_i}{r^i}  \right|_{r\gg r_c}\,.
\end{equation}
We do this for $r$ up to $10^5$.  In Fig.~\ref{asymplot}, we plot the numerical and analytical results of real and imaginary parts of the function $R$ for $r$ running from $9.9 \times 10^{4}$ to $10^5$. The solid lines are
drawn from the analytical expression \eqref{Ratlarger} and the dots  are the numerical data.
They match perfectly well at such large values of $r$, even for the exponentially growing $R$ of the stable QNM.

\subsection{Classes of unstable QNMs}

\begin{figure}[hbtp]
  \centering
  \includegraphics[width=0.48\textwidth]{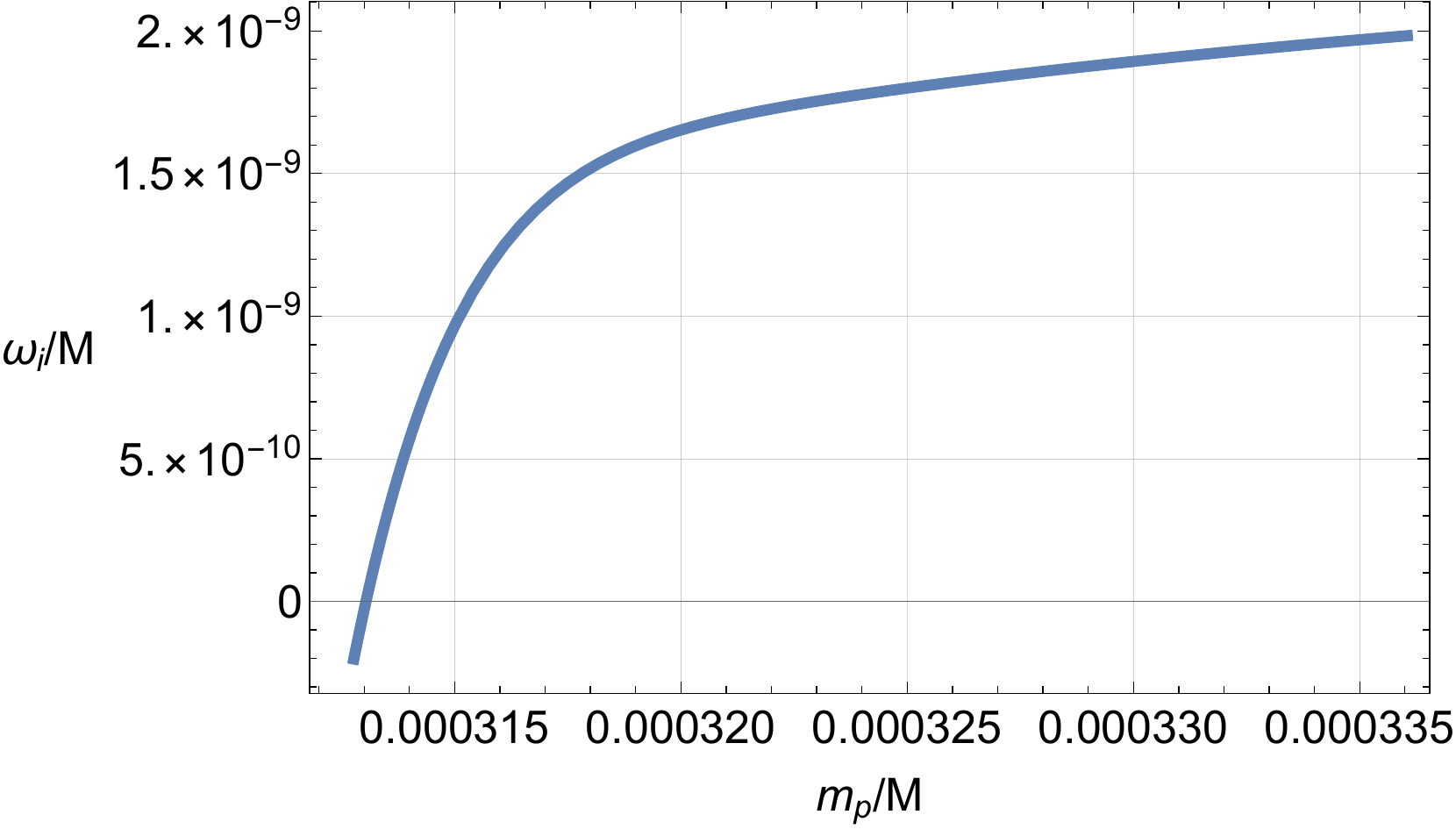}
 \caption{\small we plot the dependence of $\omega_i/M$ in terms of $m_p/M$ with fixed $(q_1,q_2,\ell,Q_1,Q_2)$=(0.025, 0.025, 1, 1, 100). There is a critical $m_p^*$, above which $\omega_i$ changes from negative to positive and the QNMs become unstable.}\label{omegai-mp}
\end{figure}

Having obtained an unstable QNM for the parameters $(m_p, q_1,q_2,\ell,Q_1,Q_2)$ =(0.045, 0.025, 0.025, 1, 1, 100), we would like to scan the parameters that allow further such modes.  We first consider varying $m_p$, with $(q_1,q_2,\ell,Q_1,Q_2)$=(0.025, 0.025, 1, 1, 100) fixed. When $m_p=0$, there can be no QBS's but only QNMs, and the effective potential has single peak only. Indeed, for sufficiently small $m_p$, the double peak potential reduces to a single peak one, and the corresponding $\omega_i$ changes the sign to become negative.  In other words, there exists a critical value of $m_p^*=0.0447$, above which the QNMs are unstable. We plot this phenomenon in Fig.~\ref{omegai-mp}, using the dimensionless parameter $\omega_i/M$ and $m_p/M$, where $M$ is the mass of the black hole, given by \eqref{masscharge}. Note that we present only the imaginary $\omega_i$ of the complex frequency. The real part $\omega_r\sim 0.0496$ has only a small deviation. This implies that for this set of parameters, $m_p$ not only has a lower bound $m_p^*$, but also has an upper bound \eqref{upmp}.

\begin{figure}[hbtp]
  \centering
  \includegraphics[width=0.45\textwidth]{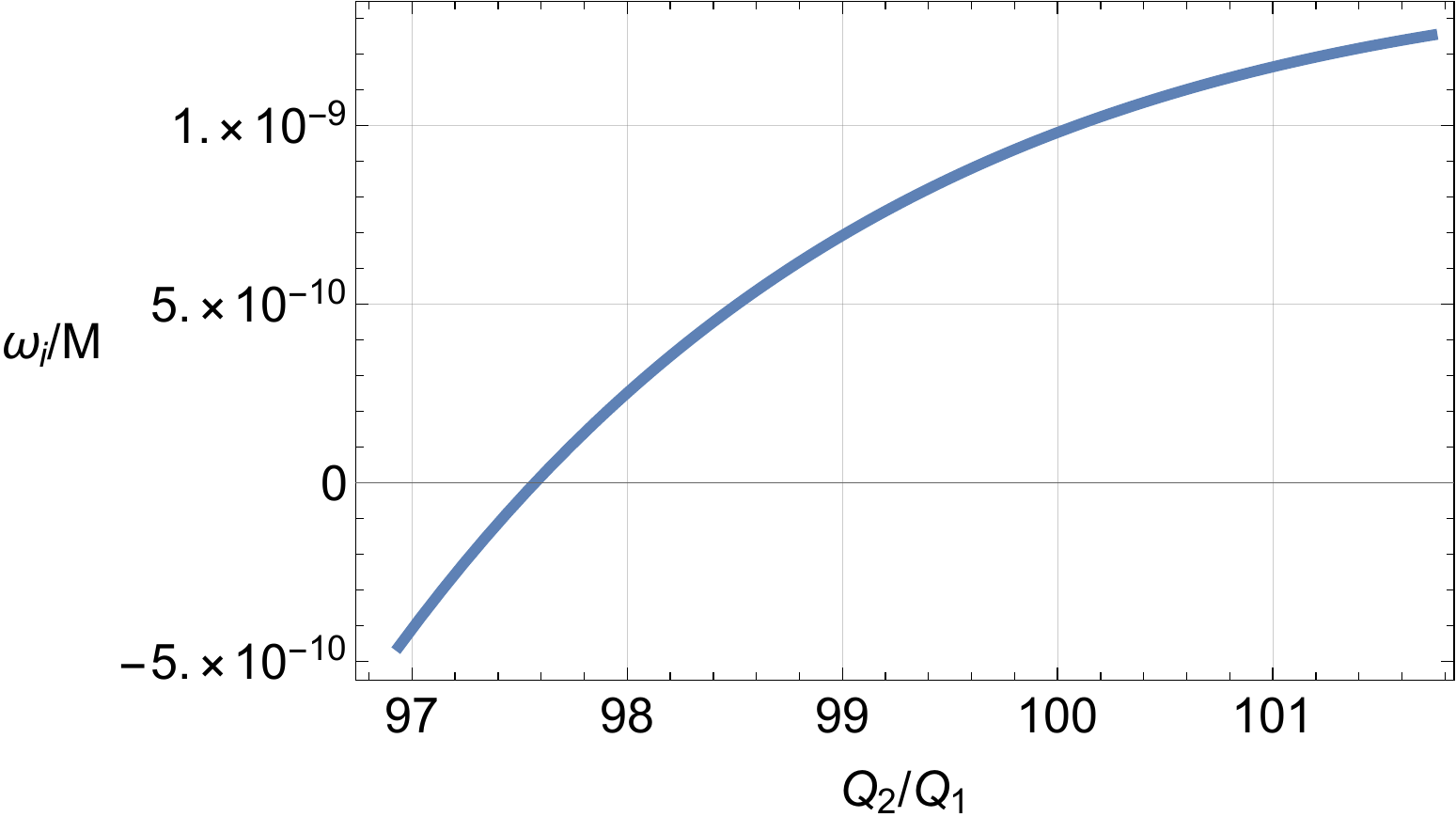}\ \ \
   \includegraphics[width=0.45\textwidth]{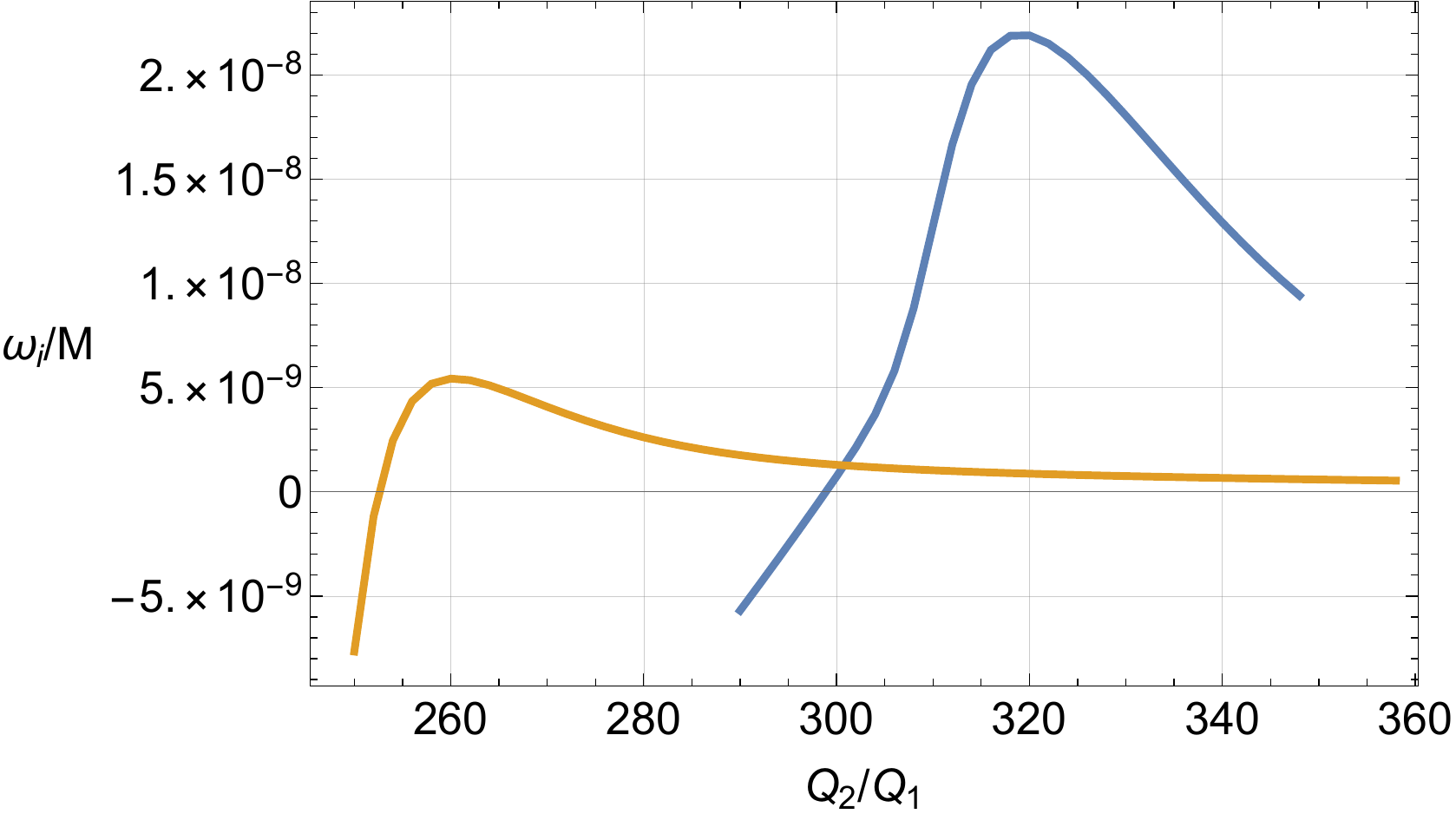}
 \caption{\small We present three branches of unstable QNMs for sufficiently large $Q_2$, with fixed $(m_p, q_1,q_2,\ell,Q_1)$ =(0.045, 0.025, 0.025, 1, 1). In each branch, there is a critical $Q_2$ beyond which unstable QNMs emerge.}\label{omegai-Q2}
\end{figure}

Next, we fix the parameters $(m_p, q_1,q_2,\ell,Q_1)$ =(0.045, 0.025, 0.025, 1, 1) and vary $Q_2$.  We find that double-peak potentials emerge only for sufficiently large $Q_2$, and this is consistent with the fact that the RN black hole has no unstable QNMs. Furthermore, there appears to be no upper limit of $Q_2$ for unstable QNMs. We further find multiple branches of QNMs, and for each branch, there is a critical value of $Q_2/Q_1$ above which unstable QNMs emerge.  We plot this phenomenon in Fig.~\ref{omegai-Q2}, using dimensionless parameter $Q_2/Q_1$. Again we present only the imaginary part of the frequency.

As we can see in Fig.~\ref{omegai-Q2}, for a suitably given black hole $(M,Q_1,Q_2)$ and fixed scalar of $(m_p,q_1,q_2)$, there can exist multiple QNMs of different frequencies. Similar occurrence happens for QBS's, which can be classified by their overtone numbers. However, how to classify and obtain the complete set of the unstable QNMs remain to be investigated.

\section{Conclusion}\label{conclu}

Motivated by the fact that black holes in STU supergravity can excite unstable QBS's of charged massive scalar fields, we constructed explicit examples of superradiantly unstable QNMs of the $(N_1,N_2)=(2,2)$ extremal black holes with the charge ratio $Q_2/Q_1$ being sufficiently large. QNMs exist naturally in black hole spacetime, satisfying the ingoing and outgoing boundary conditions on the horizon and in the asymptotic region respectively. The frequency is necessarily complex, and typically the imaginary part is negative indicating that the black hole is stable against the perturbation.  Unstable QNMs are far less common, and in fact ours are the first examples in literature. It indicates that even the static charged black holes can be vulnerable against the superradiant instability.

In order to guide our numerical construction, we studied some necessary conditions for the unstable QNMs. Intuitively, the superradiant condition must be satisfied since there is no external energy source for QNMs. We found, empirically, that effective potentials in the Schr\"odinger-like equation would have necessarily double peaks when unstable QNMs arised. In the case of QBS's, this is understandable since we need a deep-enough potential well to trap the wave to form a bound state. However, in the case of unstable QNMs, the potential well is very shallow and why such a volcano-shaped potential leads to unstable QNMs is not clear. Furthermore, the double-peak requirement appears to rule out the massless scalar fields. With our explicit construction of superradiantly unstable QNMs, the general conditions for such QNMs and how to classify them need to be thoroughly investigated.  It is also tantalizing to investigate the connections between the unstable QBS's and QNMs, and black hole echoes, which also require double-peak potentials \cite{Huang:2021qwe}.

Finally we point out that the extremal black holes we considered in this paper are superymmetric and expected to be stable in the context of supergravity. The fact that they are very vulnerable to the superradiant instability in our simple but possibly non-supersymmetric setup illustrates the need to investigate this instability properly in a fully supergravity theory.

\section*{Acknowledgement}

This work is supported in part by the National Natural Science Foundation of China (NSFC) grants No.~12005155, No.~11875200 and No.~11935009.

\providecommand{\href}[2]{#2}\begingroup\raggedright

\end{document}